\documentclass[conference]{IEEEtran}
\IEEEoverridecommandlockouts
\usepackage{cite}
\usepackage{amsmath,amssymb,amsfonts}
\usepackage{algorithmic}
\usepackage{graphicx}
\usepackage{textcomp}
\usepackage{xcolor}
\def\BibTeX{{\rm B\kern-.05em{\sc i\kern-.025em b}\kern-.08em
    T\kern-.1667em\lower.7ex\hbox{E}\kern-.125emX}}

\usepackage{caption}
\usepackage{subcaption}

\begin{document}

\title{Sound Classification of Four Insect Classes\\
}

\author{
\IEEEauthorblockN{Yinxuan Wang}
\IEEEauthorblockA{\textit{Computer and Information Technology} \\
\textit{Purdue University}\\
West Lafayette, U.S. \\
wang3910@purdue.edu}

\and

\IEEEauthorblockN{Sudip Vhaduri}
\IEEEauthorblockA{\textit{Computer and Information Technology} \\
\textit{Purdue University}\\
West Lafayette, U.S. \\
svhaduri@purdue.edu}
}

\maketitle

\begin{abstract}
The goal of this project is to classify four different insect sounds—cicada, beetle, termite, and cricket. One application of this project is for pest control to monitor and protect our ecosystem. Our project leverages data augmentation, including pitch shifting and speed changing, to improve model generalization. This project will test the performance of Decision Tree, Random Forest, SVM RBF, XGBoost, and k-NN models, combined with MFCC feature. A potential novelty of this project is that various data augmentation techniques are used and created 6 data along with the original sound. The dataset consists of the sound recordings of these four insects. This project aims to achieve a high classification accuracy and to reduce the over-fitting problem.
\end{abstract}


\section{Introduction}
Insects play critical roles in various ecosystems, influencing agriculture, forestry, and the environment at large. However, certain insect species, particularly pests, can cause significant harm to crops, infrastructure, and natural habitats. According to United States Environmental Protection Agency, termites cause damage to the houses and all kind of structures that cost about billions of dollars, and the owners of these properties have to spend more than two billion dollars to treat them \cite{usepa_termites}. According to United States Department of Agriculture, bark beetles can be harmful to environment, animals, and human beings, and they are one of the sources to cause diseases and mechanical damages, such as file, to the forests \cite{usda_bark_beetle}. Though crickets and cicadas are usually not considered as pests, they are usually very noisy in day and nights, and affect people working and resting. In a program of the United States National Institutes of Health, cicadas can make as much noise as a motorcycle \cite{nih_cicada_noisy}. According to a blog, some kinds of crickets are harmful to crops or occasionally bite humans, such as tobacco crickets and striped raspy cricket \cite{animalwised_crickets_harmful_to_human}. Hence, early and accurate identification of these insects is essential for both ecological studies and pest control efforts.

In recent years, improvements in acoustic classification have opened new possibilities for identifying insects based on their sounds. By analyzing the unique acoustic patterns produced by different insect species, models can offer a non-invasive, efficient way to monitor insect populations. The development of such models is particularly valuable for pest control, where early detection of harmful species can prevent significant economic losses.

Our project aims to contribute to this field by training classifiers which can distinguish the sounds of four specific insect species: cricket, cicada, termite, and bark beetle. To improve the generalization of the model, data augmentation, such as pitch shifting and time stretching, are used, ensuring low over-fitting to the training dataset. This project will measure and compare the performances of decision tree, random forest, k-nearest neighbor, support vector machine with radial basis function, and XGBoost, in combining with Mel-frequency Cepstral Coefficients (MFCC).

\begin{figure}
  \begin{subfigure}{0.45\columnwidth}
  \includegraphics[width=\textwidth]{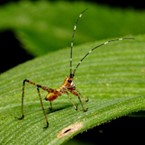}
  \caption{Cricket}
  \end{subfigure}
  \hfill
  \begin{subfigure}{0.45\columnwidth}
  \includegraphics[width=\textwidth]{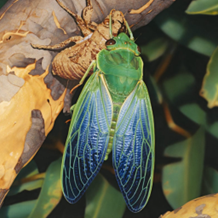}
  \caption{Cicada}
  \end{subfigure}
  \begin{subfigure}{0.45\columnwidth}
  \includegraphics[width=\textwidth]{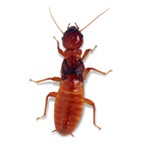}
  \caption{Termite}
  \end{subfigure}
  \hfill
  \begin{subfigure}{0.45\columnwidth}
  \includegraphics[width=\textwidth]{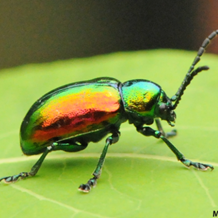}
  \caption{Beetle}
  \end{subfigure}
\caption{Figures of Four Insect Classes}
\end{figure}

\section{Related Work}

\subsection{Bag of On-Phone ANNs to Secure IoT Objects Using Wearable and Smartphone Biometrics}
S. Vhaduri, W. Cheung, and S. Dibbo explored the availability to utilize biometrics from wearables and smart phones with a bag of on-phone artificial neural network models in 2023 \cite{sudipvhaduri_2023_bag}. Although this research is not related to insect sound classification, they provided several possible and available values for pitch shifting and speed changing. In our research, we eventually selected -2, -1, 1, and 2 for pitch shifting, and 0.5x and 2.0x for speed changing.

\subsection{Insect Classification Solutions based on Insect Images}
In 2017, C. Martineau et al. tested the performance of multiple classifiers, including SVM, decision tree, MPL, Bayes, Logistic, Random Forest, k-NN, and so forth, to classify insect based insects' images \cite{martineau_2017_imagebased}. In 2018, D. Xia and et al. used an improved convolutional neural network with a multiple-kernel learning technique to detect and classify insects based on their images \cite{xia_chen_2018_insect_detection_cnn}. However, using image-based solution has some limitation. For instance, when an insect is hidden inside the house or underneath some furniture,it would be impossible to detect them via image-based solution. Hence, in our research, we explore the availability to classify insects based on their sounds, providing one more way to detect and classify insects.

\subsection{Insect Classification Solutions based on Sounds}
Researchers have been using acoustic data to identify people \cite{vhaduri2023implicit, dibbo2022phone, cheung2020continuous}, respiratory decease monitoring \cite{pfeifer2024mitigating, vhaduri2023environment, vhaduri2023transfer, dibbo2021effect}, well-being tracking \cite{vhaduri2022understanding, dibbo2021visualizing, vhaduri2021predicting, vhaduri2021deriving}, sleep health monitoring \cite{chen2020estimating, vhaduri2020nocturnal}, etc. However, acoustic data can also be used for insect classification.

X. Dong, N. Yan, and Y. Wei introduced an insect sound classifier based on convolutional neural network, and they used MFCC and chromatic spectrogram as the features in 2018 \cite{dong_2018_cnn}. Their classifier recieved an 97.8723\% accuracy rate among 47 types of insect sound from USDA library, which is also one of the datasets our research will use \cite{dong_2018_cnn}. However, they trained the classifiers using the images of the spectrogram as the dataset. Similarily, M. Zhang et al. also used CNN and MFCC features to train the classifier based on the images of the spectrogram, yet they only received 92.56\% accuracy among nine insect species in 2021 \cite{zhang_2021_mfcc_cnn}. In comparison to their methodology, we explored the availability of training classifiers with the coefficients of MFCC features rather than the images of the spectrogram.

In 2022, S. Basak et al. explored the accuracy and limitation of various k-NN and SVM classifiers with MFCC feature \cite{basak_2022_knn}. Surprisingly, their accuracy is relatively lower than other research work and our result: the accuracy of k-NN, cosine k-NN, medium k-NN, SVM, and Linear SVM are 85.4\%, 83.9\%, and 83.8\%, 84.6\%, and 84.0\%, respectively. In our research, we also test the performance of k-NN as well as SVM classifiers, and we achieve a higher accuracy than their result. A possible reason could be that we extract 40 MFCC features rather than only 13 MFCC features, and besides, we applied data augmentation after segmentation to increase the size of the dataset.

\subsection{Other Insect Classification Solutions}
Y. Chen et al. designed a new inexpensive sensor and used a Bayesian classifier to classify flying insects by the incidental sound of their flight in 2014 \cite{chen_2014_flying_insect_classification}. Their solution did solve the issue that the traditional acoustic method is difficult to gather a large dataset and thus could result into poor accuracy and classifier performance. In contrast, our research uses data augmentation to increase the size of the dataset. We also test the performance of more machine learning classifiers that can be later comparing with the result in their work.

\subsection{Summary}
Although all of these related work have some contributions on insect sound classification, currently no published research paper works on comparing five different models with 40 MFCC features, and only trained through the coefficients rather than images of the spectrograms. Hence, our research will contribute to this area, especially classifying the four sound insect species.

\section{Methodologies}
We will implement random forest, decision tree, k-NN, SVM RBF, and XGBoost models and extract 40 MFCC features for training and testing. Each model will be trained and tested separately.

Decision Tree is a very common classifier, and its result is usually easy to interpret.

Random Forest is also a widely used model for classification and it often has higher accuracy and can prevent overfitting problems compared to a single decision tree.

MFCC is a popular feature in speech recognition and audio signal processing because it can keep important acoustic information to make the model easier to classify.

The code of this project is stored in Purdue GitHub: \\
https://github.itap.purdue.edu/wang3910/CNIT581-Practicum-1.

\section{Dataset}
Our dataset consists of the sound recordings from four insect categories: cicada, beetle, termite, and cricket. The dataset is sourced from other research as listed after the figures below. Our dataset will be divided, and 80\% are the training set while 20\% are the test set. In the training set, the audio recordings will be augmented using pitch shifting and speed changing.

\subsection{ESC-50}
ESC-50, which is the dataset for environmental sound classification, consists of 2000 labeled environmental audio recordings in total, and we need the flying insect recordings for this project \cite{piczak_2015_esc}. This dataset has 2000 environmental audio recordings, and it consists of 50 semantical classes in five major categories: animals, natural and water sounds, human non-speech sounds, domestic sounds, and urban noises \cite{piczak_2015_esc}.

\subsection{InsectSet32: Datasets for Automatic Acoustic Identification of Insects (Orthoptera and Cicadidae)}
This dataset has 335 audio recordings of 32 sound producing insect species: 147 recordings are belonging to nine species from the order Orthoptera, while the rest 188 recordings are belong to 23 species in the family Cicada \cite{fai_2022_insectset32}. This project will use these 188 recordings that are belonging to Cicadidae family.

\subsection{Bug Bytes Sound Library: Stored Product Insect Pest Sounds}
Bug Bytes Sound Library is provided by the ARS Center for Medical, Agricultural, and Veterinary Entomology, and is used to support the detection and control of hidden insect infestations \cite{mankin_2019_bug}. This dataset consists of 52 species that are considered as pests. This project will use five different termite species from the dataset.

\subsection{Experimental Characterization and Automatic Identification of Stridulatory Sounds Inside Wood – Supplementary Information (Data)}
This dataset is the supplementary information for the research conducted by Carol L. Bedoya, Ximena J. Nelson, Eckehard G. Brockerhoff, Stephen Pawson, and Michael Hayes, and consists of 360 acoustics signals from Hylurgus ligniperda and Hylastes ater, which are two beetle species \cite{bedoya_2022_experimental}.

\subsection{Data Segmentation}
\begin{figure}[h]
\centerline{\includegraphics[width=0.5\textwidth]{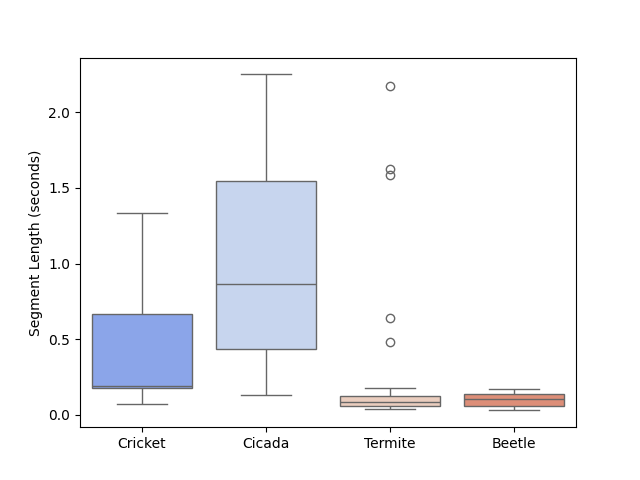}}
\caption{The box plot of segment durations in each class.}
\label{plot_segment_duration}
\end{figure}

This project will use Audacity, a free audio editing app, to do audio clip segmentation. We only pick the sounds from four insect classes and ignore other ones, such as birdsong. To easier displaying and future processing, each insect class has a corresponding id: C1 refers to crickets, C2 refers to cicadas, C3 refers to termites, and C4 refers to bark beetles. As shown in figure \ref{plot_segment_duration}, most segments are less than three seconds long, though they have several outliers. Yet, class C3 have multiple segments which are about 10 seconds long. Among 25 original clips we choose from five classes correspondingly, class C4 has 382 segments, which is the most, and class C3 has 17 segments, which is the least. The minimum segment duration ($w$) is approximately 0.0288 seconds, and the minimum sample size is 636.

Then, we create instances from all segments based on $w$. The segments that are longer than $w$ will be divided into multiple instances. The last instance can be overlapped with its previous instance because a segment's duration is not necessarily an integral multiple of $w$.

\begin{figure}[h]
\centerline{\includegraphics[width=0.5\textwidth]{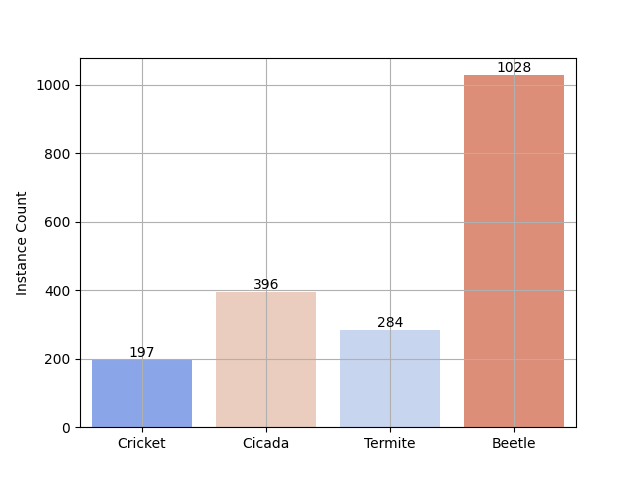}}
\caption{The bar plot of instances in each class.}
\label{plot_balanced_instances}
\end{figure}

As shown in figure \ref{plot_balanced_instances}, class cricket has the least number of instances. Hence, we will randomly choose 663 instances from each class to apply data augmentation in the next step.

\subsection{Data Augmentation}
Pitch shifting and speed changing are the data augmentation techniques we are going to use in this project. The ranges for pitch shifting and speed changing come from Professor Sudip Vhaduri, W. Cheung, and S. V. Dibbo’s research \cite{sudipvhaduri_2023_bag}.

In pitch shifting, the values of the pitches are ranged from -3.5 to 3.5 with 0.5 increments. Thus, there are 14 variations of different pitches, excluding the original pitch.

In time stretching, the values of the speeds are ranged from 0.25x to 2x with 0.25 increments. Thus, there are seven variations of different speeds, excluding the original speed.

Hence, each audio recording in the training set becomes 22 augmented data, which have one original recording plus 14 pitch variations plus seven speed variations.

After data augmentation, each class has exactly 14586 augmented class balanced instances which are ready for training and evaluating the models.

\subsection{Dataset Naming Convention}
Since our dataset is combined with multiple datasets from other research, we decide to keep the original file names, and append extra information after the file names, so the convention becomes: \{Original File Name\}\#\{Segment Number\}\#\{$w$Number\}\#\{Augmentation Type\}\#1.wav. The augmentation type is $P$, pitch shifting, or $T$, time stretching.

\section{Results and Evaluations}
This project will test and evaluate the performance of Decision Tree, Random Forest and k-NN classifiers based on accuracy and confusion matrix with different segment lengths, different balanced instance count, and whether dataset will be augmented.

\subsection{t-SNE and UMAP Plots}
\begin{figure}[h]
    \centering
    \includegraphics[width=0.45\textwidth]{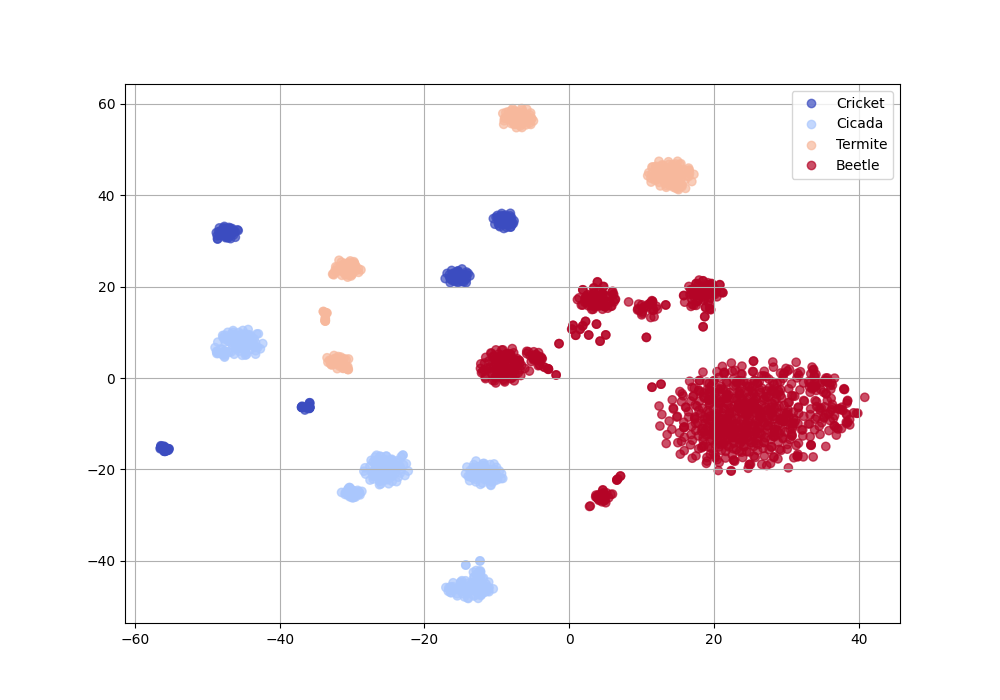}
    \caption{t-SNE plot of instances in four classes.}
    \label{fig:tsne_mfcc40}
\end{figure}

\begin{figure}[h]
    \centering
    \includegraphics[width=0.45\textwidth]{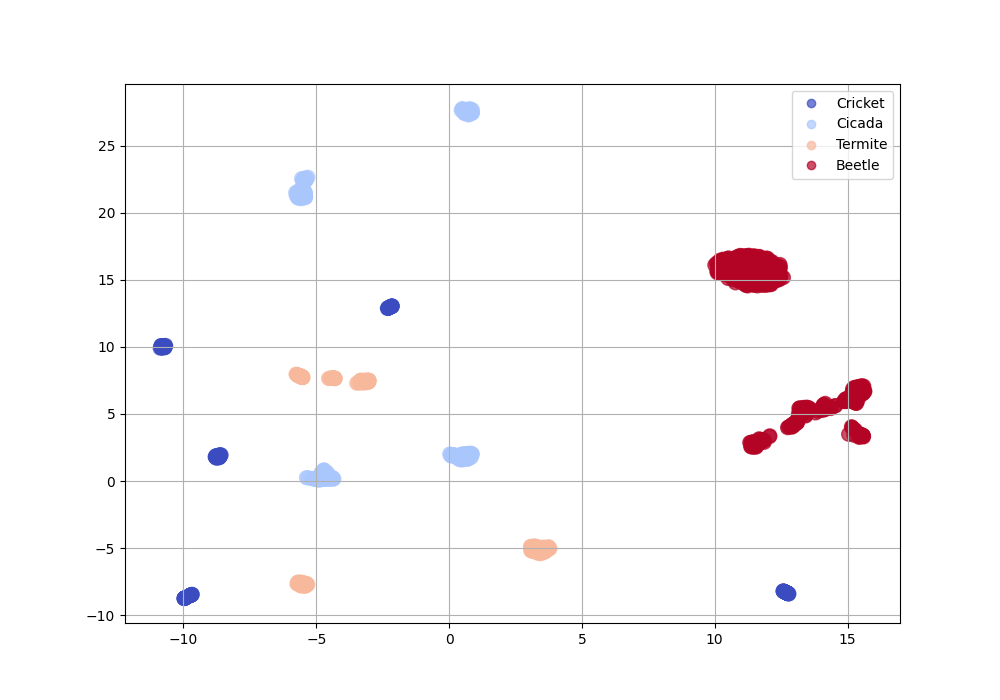}
    \caption{UMAP plot of instances in four classes.}
    \label{fig:umap_mfcc40}
\end{figure}

\begin{figure}[h]
    \centering
    \includegraphics[width=0.45\textwidth]{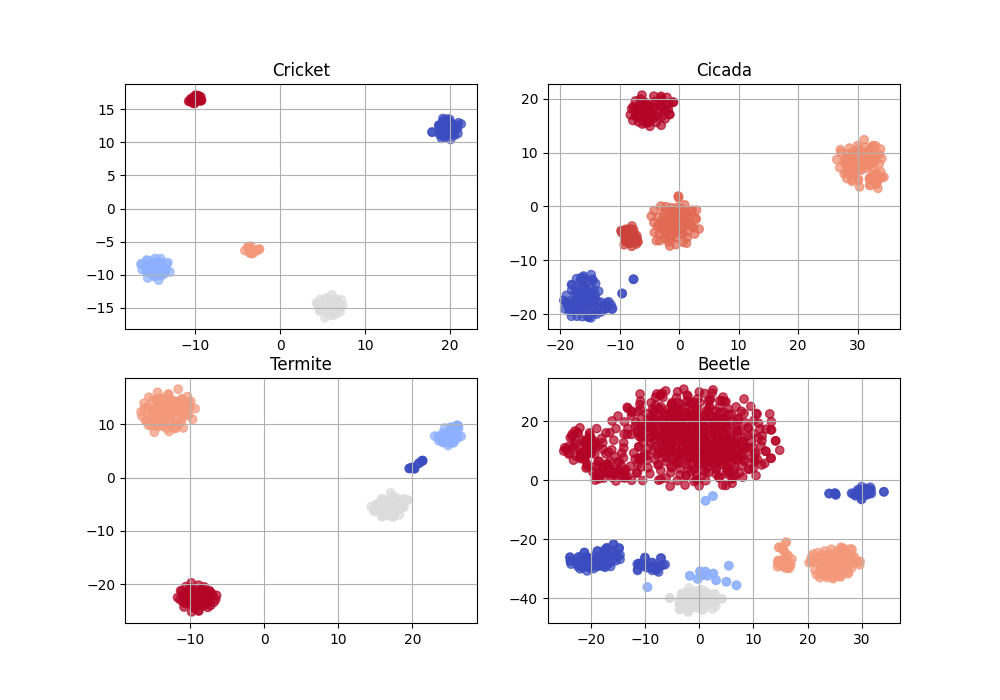}
    \caption{t-SNE plot of instances grouped by their original clips in each class.}
    \label{fig:tsne_mfcc40_originalclips}
\end{figure}

\begin{figure}[h]
    \centering
    \includegraphics[width=0.45\textwidth]{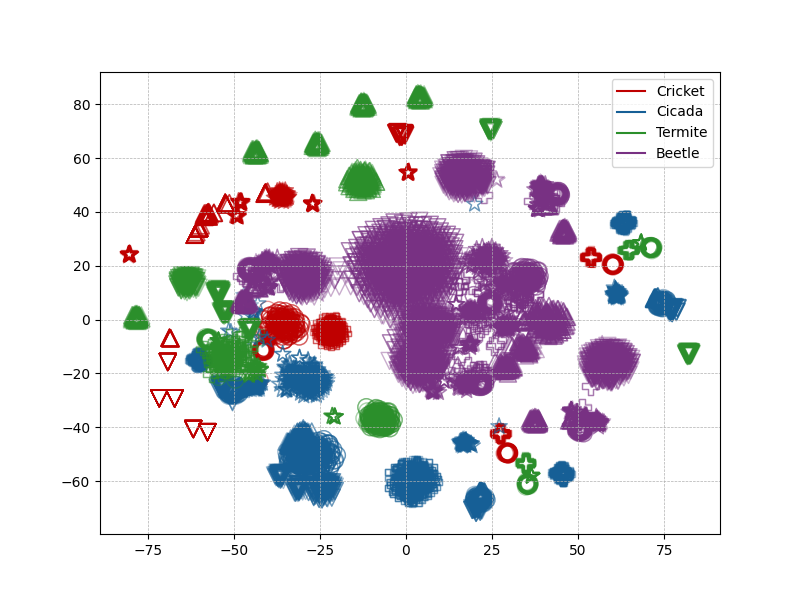}
    \caption{t-SNE plot of instances grouped by their original clips in each class after augmentation.}
    \label{fig:tsne_mfcc40_originalclips_allsep_aug}
\end{figure}

The results of both t-SNE and UMAP plots are visualized in Figures \ref{fig:tsne_mfcc40} and \ref{fig:umap_mfcc40}. These methods are used to project 40 MFCC features into a 2D space to easily observe patterns among four classes. As shown in both graph, the beetles are more clustered closely, while the rest three classes are more sparse. Hence, we predicted that the trained classifier would have a better performance on beetles, but not on crickets, cicada, and termites.
In Figure \ref{fig:tsne_mfcc40_originalclips}, you can see that the beetles have closer distances among five original clips, while some original clips from cricket, cicada, and termite classes are further apart. Hence, we could conclude that beetles will have better results compared to the other three classes.
Figure \ref{fig:tsne_mfcc40_originalclips_allsep_aug} demonstrates the t-SNE plot of all instances after data augmentation. In the plot, many instances in different classes are overlapping, indicating the difficulties of models to classify these data.

\begin{figure}[h]
    \centering
    \includegraphics[width=0.45\textwidth]{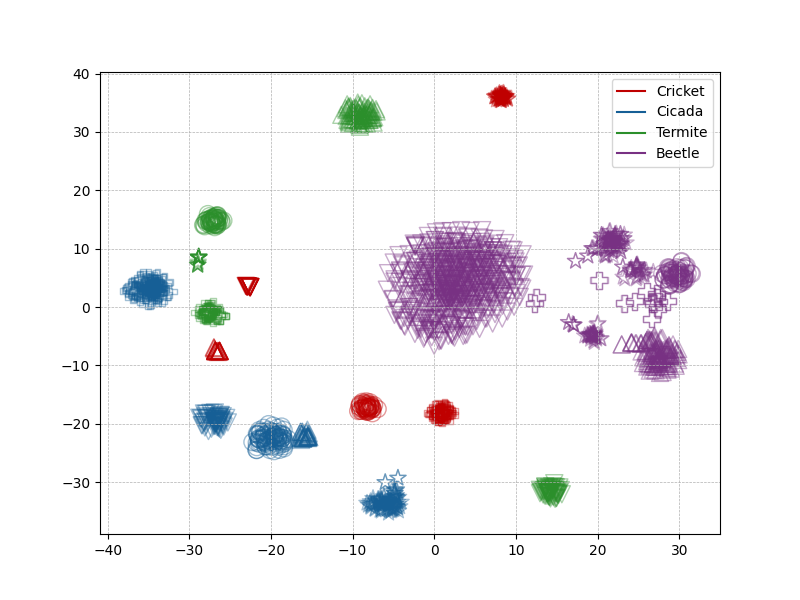}
    \caption{t-SNE plot of instances grouped by their original clips in each class before augmentation.}
    \label{fig:tsne_mfcc40_originalclips_allsep}
\end{figure}

\subsection{Results with Segment Length 0.1 Without Separating Original Clips}
We first use 0.1 as the segment length, and cricket has 197 instances, cicada has 396 instances, termite has 284 instances, and beetle has 1028 instances, as shown in \ref{plot_balanced_instances}. So, we choose 30 and 150 for each trial, and we will use decision tree, random forest, and k-NN models with MFCC features.

\subsubsection{Results of Balanced Instance Count 30}
All models perform well when using 30 as a balanced instance count, and as shown in all confusion matrix plots, the Decision Tree (DT), Random Forest (RF), and k Nearest Neighbor (k-NN) classifiers can distinguish between four insect sounds: Cricket, Cicada, Termite, and Beetle in very high accuracy. In these matrices, each row represents the true labels, and each column represents the predicted labels.

\subsubsection{Results of Balanced Instance Count 150}
All classifiers also perform well when using 150 as balanced instance count, and as shown in all confusion matrix plots, that Decision Tree, Random Forest, and k-NN classifiers, can distinguish between four insect sounds: Cricket, Cicada, Termite, and Beetle in a very high accuracy.

\subsubsection{Summary}
These results have very high accuracies might due to the problem that we did not separate the original clips between training set and test set. Thus, all classifiers are over-fitting to the given training set and test set. We predict that they would not do well on the real-world samples, due to the over-fitting problem. Hence, we will give a second try with the same segment length and, at the same time, separating original clips.

\subsection{Results of Segment Length 0.1 with Separating Original Clips}
In the second try, we still use 0.1 as the segment length and Decision Tree as the classifier, but the process is different. In each class, the unbalanced instances are grouped by their original clips, and then instances from one original clip work as the test set, and we balance the instances from the other four original clips, which is the training set.

\begin{table}[h!]
    \centering
    \refstepcounter{table}
    \textbf{\large Table \thetable: Instance Counts from Original Clips in Leave-1-out Tests} \\[1ex]
    \begin{tabular}{c|c|c|c|c|c}
        \hline
        Training Set & Test Set & Cricket & Cicada & Termite & Beetle \\ [0.5ex]
        \hline
        1, 2, 3, 4 & 5 & 167 & 321 & 217 & 394 \\
        1, 2, 3, 5 & 4 & 172 & 361 & 184 & 876 \\
        1, 2, 4, 5 & 3 & 150 & 305 & 227 & 892 \\
        1, 3, 4, 5 & 2 & 149 & 297 & 240 & 950 \\
        2, 3, 4, 5 & 1 & 150 & 300 & 268 & 1000 \\
        \hline
    \end{tabular}
    \caption*{The instance counts from each original clips across the four classes.}
    \label{table:i=0.1_instance_counts_from_original_clips}
\end{table}

\subsubsection{Results of Balanced Instance Count 30}
Interestingly, higher testing accuracy occurs in MFCC-40 and its top 10 features, while top 20 and top 30 have lower accuracy.
The five tests give different top 10, 20, and 30 MFCC features. They do have some common features: 0, 1, 2, 3, and 4. These five features appear throughout all tests. Features 7, 42, 86, 87, 88, and 97 are common in all top 30 MFCC features.

\begin{table}[h!]
    \centering
    \refstepcounter{table}
    \textbf{\large Table \thetable: Sum of Importance Scores} \\[1ex]
    \begin{tabular}{c|c|c|c|c}
        \hline
        Training Set & Test Set & Top 10 (\%) & Top 20 (\%) & Top 30 (\%) \\ [0.5ex]
        \hline
        1, 2, 3, 4 & 5 & 20.637 & 32.922 & 43.565 \\
        1, 2, 3, 5 & 4 & 39.153 & 50.238 & 58.458 \\
        1, 2, 4, 5 & 3 & 51.769 & 58.789 & 64.889 \\
        1, 3, 4, 5 & 2 & 39.687 & 49.263 & 57.117 \\
        2, 3, 4, 5 & 1 & 43.963 & 51.934 & 58.757 \\
        \hline
    \end{tabular}
    \caption*{The sums of importance scores for top-selected MFCC features with 30 instances.}
    \label{table:sum_importance_scores_i30}
\end{table}

\begin{figure}[h]
    \centering
    \includegraphics[width=0.45\textwidth]{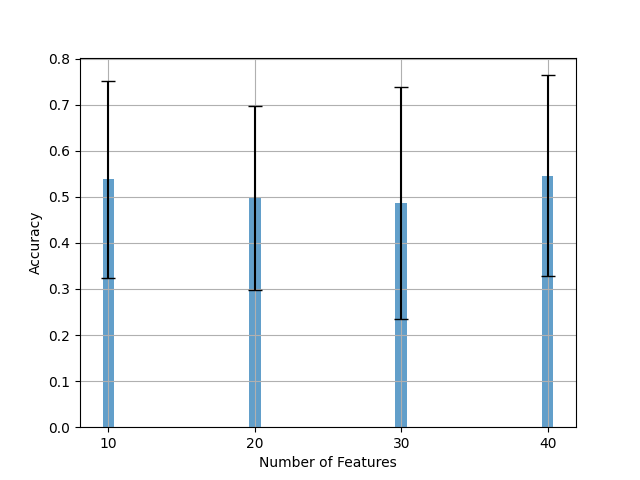}
    \caption{Bar Plot of Decision Tree (i=30).}
    \label{fig:i30_dt_mfcc40_233873162}
\end{figure}

\subsubsection{Results of Balanced Instance Count 145}
We choose 145 as the balanced instance count because the minimum number of instances is 149 as shown in Table \ref{table:i=0.1_instance_counts_from_original_clips}. Similar to the result of 30 balanced instances, the classifier with full MFCC-40 features still have the highest testing accuracy. They also do not perform well when picking the top 20 and 30 MFCC features.

\begin{figure}[h]
    \centering
    \includegraphics[width=0.45\textwidth]{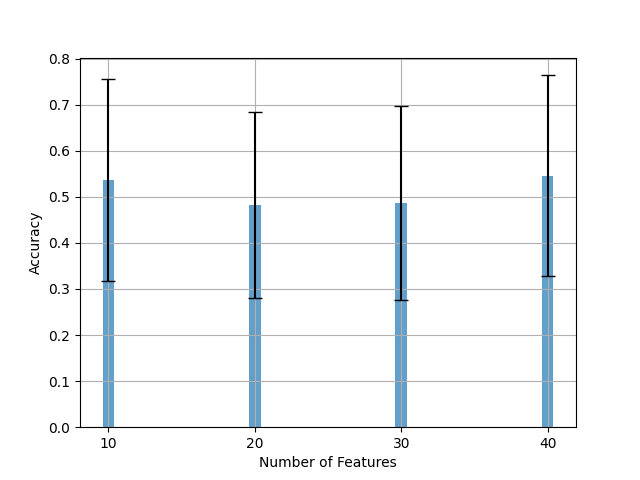}
    \caption{Bar Plot of Decision Tree (i=145).}
    \label{fig:i145_dt_mfcc40_233873162}
\end{figure}

\begin{table}[h!]
    \centering
    \refstepcounter{table}
    \textbf{\large Table \thetable: Sum of Importance Scores} \\[1ex]
    \begin{tabular}{c|c|c|c|c}
        \hline
        Training Set & Test Set & Top 10 (\%) & Top 20 (\%) & Top 30 (\%) \\ [0.5ex]
        \hline
        1, 2, 3, 4 & 5 & 20.637 & 32.922 & 43.565 \\
        1, 2, 3, 5 & 4 & 39.153 & 50.238 & 58.458 \\
        1, 2, 4, 5 & 3 & 51.769 & 58.789 & 64.889 \\
        1, 3, 4, 5 & 2 & 39.687 & 49.263 & 57.117 \\
        2, 3, 4, 5 & 1 & 43.963 & 51.934 & 58.757 \\
        \hline
    \end{tabular}
    \caption*{The sums of importance scores for top-selected MFCC features with 145 instances.}
    \label{table:sum_importance_scores_i145}
\end{table}

\subsection{Results of Segment Length 0.1 with Separating Original Clips and Augmentation}
In the third try, we still use 0.1 as the segment length, and we trained more classifiers: decision tree, random forest, k-NN, SVM RBF, and XGBoost. The procedure is similar to the previous try, but each clip is augmented and padded into same length, and we decided to use two ways: (1) pitch shift and (2) speed change.

In pitch shifting, the values of the pitches are ranged from -2 to 2 with 1.0 increment. Thus, there are four variations of different pitches, excluding the original pitch.

In time stretching, the values of the speeds are 0.5x and 2.0x. Thus, there are two variations of different speeds, excluding the original speed.

Hence, each audio recording in the training set becomes seven augmented data, which have one original recording plus six pitch variations plus seven speed variations.

\subsubsection{Decision Tree with 30 Instances}
\begin{figure*}
  \begin{subfigure}{0.45\columnwidth}
  \includegraphics[width=\textwidth]{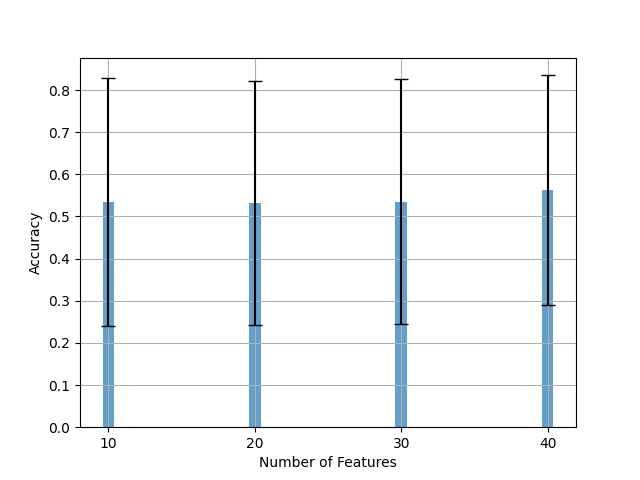}
  \caption{Before augmentation.}
  \end{subfigure}
  \hfill
  \begin{subfigure}{0.45\columnwidth}
  \includegraphics[width=\textwidth]{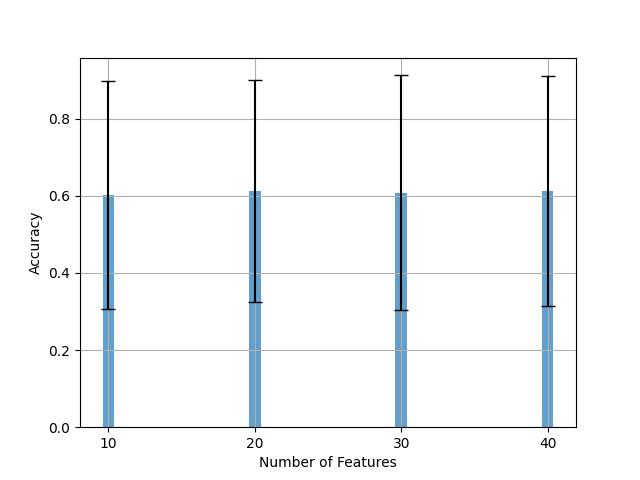}
  \caption{After augmentation.}
  \end{subfigure}
\caption{Test Accuracies of Decision Tree (i=30)}
\label{fig:i30_dt_mfcc40_acc}
\end{figure*}

As shown in table \ref{table:i30_dt_mfcc40_acc} and figure \ref{fig:i30_dt_mfcc40_acc}, the average accuracies of top 10, 20, 30, and 40 MFCC features are increased by about 6.790\%, 8.048\%, 7.281\%, and 7.901\% after data augmentation, respectively. Among the training sets, the set 1, 2, 3, and 4 achieved consistently high accuracies, with minimal fluctuations, peaking at 96.464\% for the top 20 features. Sets like 1, 2, 4, 5 and 1, 3, 4, 5 showed remarkable improvement post-augmentation, nearly doubling in some cases (e.g., 27.795\% to 53.890\% for 1, 2, 4, 5 with the top 10 features). However, the set 2, 3, 4, 5 displayed a significant decrease in performance after augmentation, suggesting potential issues such as noise or data imbalance.

\begin{table}[h!]
    \centering
    \refstepcounter{table}
    \textbf{\large Table \thetable: Test Accuracies of Decision Tree (i=30)} \\[1ex]
    \begin{tabular}{c|c|c|c}
        \hline
        Training & Top & Before & After \\ [0.5ex]
        Set & Features & Aug. (\%) & Aug. (\%) \\ [0.5ex]
        \hline
        Average & 10 & 53.432 & 60.222 \\
        Average & 20 & 53.206 & 61.254 \\
        Average & 30 & 53.547 & 60.828 \\
        Average & 40 & 53.432 & 61.333 \\
        1,2,3,4 & 10 & 95.254 & 96.262 \\
        1,2,3,4 & 20 & 94.122 & 96.464 \\
        1,2,3,4 & 30 & 94.851 & 95.285 \\
        1,2,3,4 & 40 & 96.262 & 95.565 \\
        1,2,3,5 & 10 & 71.474 & 83.494 \\
        1,2,3,5 & 20 & 71.474 & 84.375 \\
        1,2,3,5 & 30 & 71.474 & 85.597 \\
        1,2,3,5 & 40 & 71.474 & 84.896 \\
        1,2,4,5 & 10 & 27.795 & 53.890 \\
        1,2,4,5 & 20 & 27.795 & 52.492 \\
        1,2,4,5 & 30 & 28.399 & 54.154 \\
        1,2,4,5 & 40 & 40.181 & 53.248 \\
        1,3,4,5 & 10 & 28.253 & 46.840 \\
        1,3,4,5 & 20 & 28.253 & 51.255 \\
        1,3,4,5 & 30 & 28.625 & 46.236 \\
        1,3,4,5 & 40 & 28.996 & 51.069 \\
        2,3,4,5 & 10 & 44.385 & 21.324 \\
        2,3,4,5 & 20 & 44.385 & 22.861 \\
        2,3,4,5 & 30 & 44.385 & 21.390 \\
        2,3,4,5 & 40 & 44.385 & 21.190 \\
        \hline
    \end{tabular}
    \caption*{The comparison of the test accuracies before and after data augmentation.}
    \label{table:i30_dt_mfcc40_acc}
\end{table}

\subsubsection{Decision Tree with 145 Instances}
\begin{figure}
  \begin{subfigure}{0.45\columnwidth}
  \includegraphics[width=\textwidth]{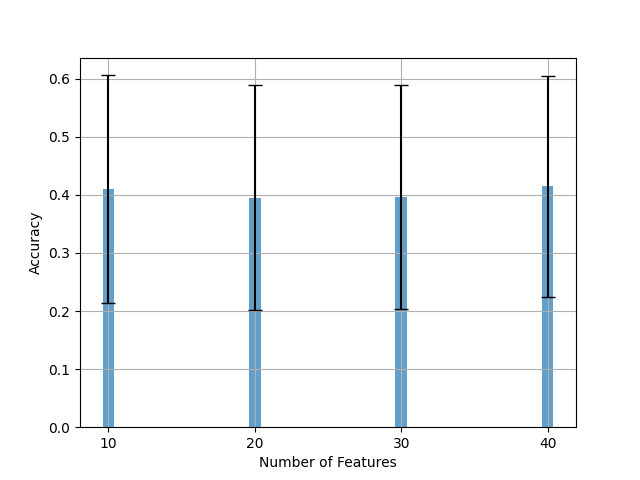}
    \caption{Before augmentation.}
  \end{subfigure}
  \hfill
  \begin{subfigure}{0.45\columnwidth}
  \includegraphics[width=\textwidth]{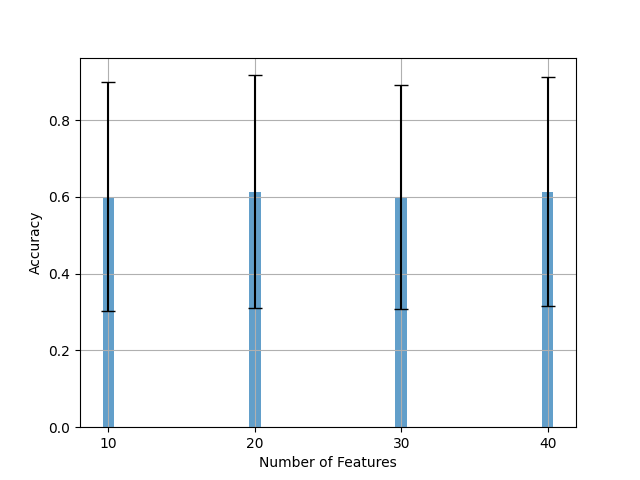}
    \caption{After augmentation.}
  \end{subfigure}
\caption{Test Accuracies of Decision Tree (i=145)}
\label{fig:i145_dt_mfcc40_acc}
\end{figure}

As shown in table \ref{table:i145_dt_mfcc40_acc} and figure \ref{fig:i145_dt_mfcc40_acc}, all average accuracies are dramatically improved after data augmentation. The average accuracies of top 10, 20, 30, and 40 MFCC features are increased by 19.010\%, 21.858\%, 20.317\%, and 19.884\% after data augmentation, respectively.
The set 1, 2, 3, and 4 exhibited the most dramatic improvement, with accuracies reaching as high as 96.262\% for the top 40 features after augmentation, compared to only 22.208\% prior. Similarly, the set 1, 2, 3, and 5 showed substantial gains, with post-augmentation accuracies peaking at 90.665\% for the top 20 features. In contrast, the set 2, 3, 4, and 5 demonstrated a performance drop after augmentation, with accuracies decreasing from 44.385\% to as low as 21.257\%, possibly due to data imbalance or overlapping features in the augmented dataset.

\begin{table}[h!]
    \centering
    \refstepcounter{table}
    \textbf{\large Table \thetable: Test Accuracies of Decision Tree (i=145)} \\[1ex]
    \begin{tabular}{c|c|c|c}
        \hline
        Training & Top & Before & After \\ [0.5ex]
        Set & Features & Aug. (\%) & Aug. (\%) \\ [0.5ex]
        \hline
        Average & 10 & 41.024 & 60.034 \\
        Average & 20 & 39.539 & 61.397 \\
        Average & 30 & 39.614 & 59.931 \\
        Average & 40 & 41.449 & 61.333 \\
        1,2,3,4 & 10 & 20.223 & 94.944 \\
        1,2,3,4 & 20 & 24.442 & 93.734 \\
        1,2,3,4 & 30 & 25.558 & 95.053 \\
        1,2,3,4 & 40 & 22.208 & 96.262 \\
        1,2,3,5 & 10 & 71.474 & 84.255 \\
        1,2,3,5 & 20 & 71.474 & 90.665 \\
        1,2,3,5 & 30 & 71.474 & 81.811 \\
        1,2,3,5 & 40 & 71.474 & 84.896 \\
        1,2,4,5 & 10 & 40.785 & 53.852 \\
        1,2,4,5 & 20 & 28.399 & 53.210 \\
        1,2,4,5 & 30 & 28.399 & 53.248 \\
        1,2,4,5 & 40 & 40.181 & 53.248 \\
        1,3,4,5 & 10 & 28.253 & 45.864 \\
        1,3,4,5 & 20 & 28.996 & 46.515 \\
        1,3,4,5 & 30 & 28.253 & 48.420 \\
        1,3,4,5 & 40 & 28.996 & 51.069 \\
        2,3,4,5 & 10 & 44.385 & 21.257 \\
        2,3,4,5 & 20 & 44.385 & 22.861 \\
        2,3,4,5 & 30 & 44.385 & 21.123 \\
        2,3,4,5 & 40 & 44.385 & 21.190 \\
        \hline
    \end{tabular}
    \caption*{The comparison of the test accuracies before and after data augmentation.}
    \label{table:i145_dt_mfcc40_acc}
\end{table}

\subsubsection{Random Forest with 30 Instances}
\begin{figure}
  \begin{subfigure}{0.45\columnwidth}
  \includegraphics[width=\textwidth]{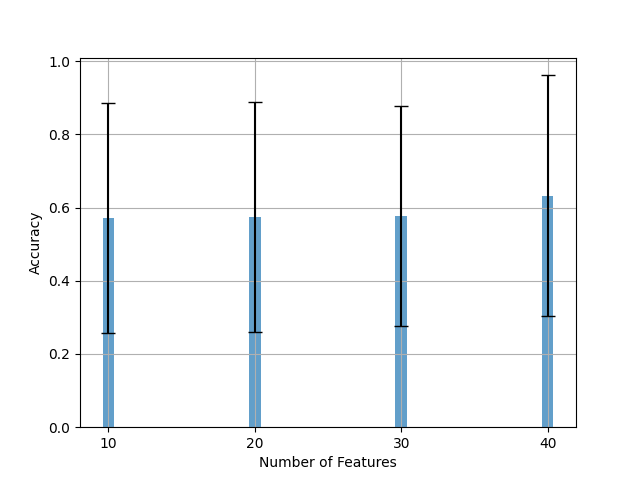}
  \caption{Before augmentation.}
  \end{subfigure}
  \hfill
  \begin{subfigure}{0.45\columnwidth}
  \includegraphics[width=\textwidth]{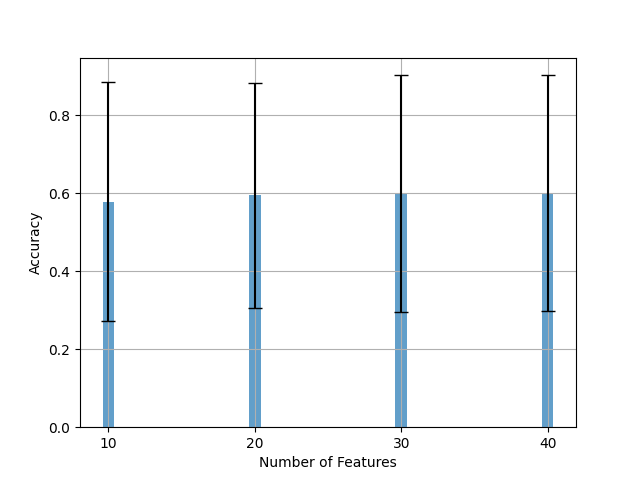}
  \caption{After augmentation.}
  \end{subfigure}
\caption{Test Accuracies of Random Forest (i=30)}
\label{fig:i30_rf_mfcc40_acc}
\end{figure}

As shown in table \ref{table:i30_rf_mfcc40_acc} and figure \ref{fig:i30_rf_mfcc40_acc}, the average accuracies of top 10, 20, and 30 MFCC features are slightly improved after data augmentation. The average accuracy of all 40 MFCC features even decreased by 3.118\% after data augmentation.
The set 1, 2, 3, and 4 consistently demonstrated high performance, maintaining accuracy levels above 96\% across all feature subsets, with minimal impact from augmentation. However, other training set combinations exhibited mixed results. For instance, the set 1, 3, 4, 5 improved significantly after augmentation, especially for the top 20 and 30 features (e.g., 30.112\% to 45.260\% for the top 20 features). Conversely, the accuracies of set 2, 3, 4, 5 declines consistently in all feature subsets, potentially due to overlap or noise introduced during data augmentation.

\begin{table}[h!]
    \centering
    \refstepcounter{table}
    \textbf{\large Table \thetable: Test Accuracies of Random Forest (RF) (i=30)} \\[1ex]
    \begin{tabular}{c|c|c|c}
        \hline
        Training & Top & Before & After \\ [0.5ex]
        Set & Features & Aug. (\%) & Aug. (\%) \\ [0.5ex]
        \hline
        1,2,3,4 & 10 & 97.270 & 97.658 \\
        1,2,3,4 & 20 & 98.139 & 97.860 \\
        1,2,3,4 & 30 & 96.402 & 96.790 \\
        1,2,3,4 & 40 & 98.015 & 98.061 \\
        1,2,3,5 & 10 & 80.128 & 78.886 \\
        1,2,3,5 & 20 & 79.487 & 77.123 \\
        1,2,3,5 & 30 & 78.846 & 83.053 \\
        1,2,3,5 & 40 & 94.231 & 81.050 \\
        1,2,4,5 & 10 & 54.381 & 54.381 \\
        1,2,4,5 & 20 & 54.381 & 54.381 \\
        1,2,4,5 & 30 & 54.381 & 54.381 \\
        1,2,4,5 & 40 & 61.329 & 54.909 \\
        1,3,4,5 & 10 & 28.996 & 35.874 \\
        1,3,4,5 & 20 & 30.112 & 45.260 \\
        1,3,4,5 & 30 & 32.714 & 44.935 \\
        1,3,4,5 & 40 & 39.033 & 45.539 \\
        2,3,4,5 & 10 & 25.134 & 22.794 \\
        2,3,4,5 & 20 & 25.134 & 22.727 \\
        2,3,4,5 & 30 & 25.668 & 20.655 \\
        2,3,4,5 & 40 & 23.529 & 20.989 \\ 
        \hline
        Average & 10 & 57.182 & 57.919 \\
        Average & 20 & 57.450 & 59.470 \\
        Average & 30 & 57.602 & 59.963 \\
        Average & 40 & 63.228 & 60.110 \\
        \hline
    \end{tabular}
    \caption*{The comparison of the test accuracies before and after data augmentation.}
    \label{table:i30_rf_mfcc40_acc}
\end{table}

\subsubsection{Random Forest with 145 Instances}
\begin{figure}
  \begin{subfigure}{0.45\columnwidth}
  \includegraphics[width=\textwidth]{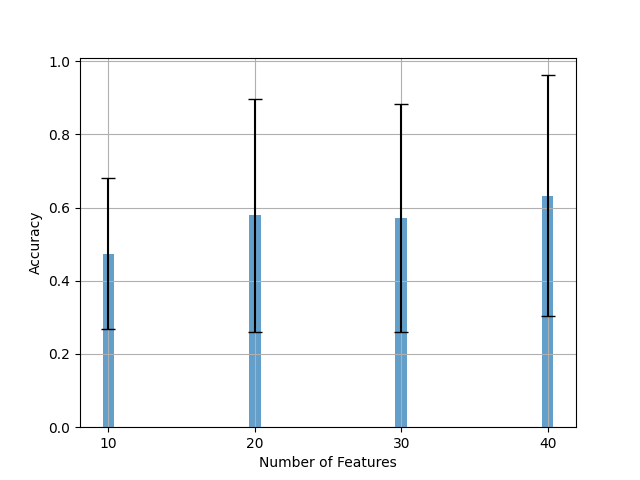}
  \caption{Before augmentation.}
  \end{subfigure}
  \hfill
  \begin{subfigure}{0.45\columnwidth}
  \includegraphics[width=\textwidth]{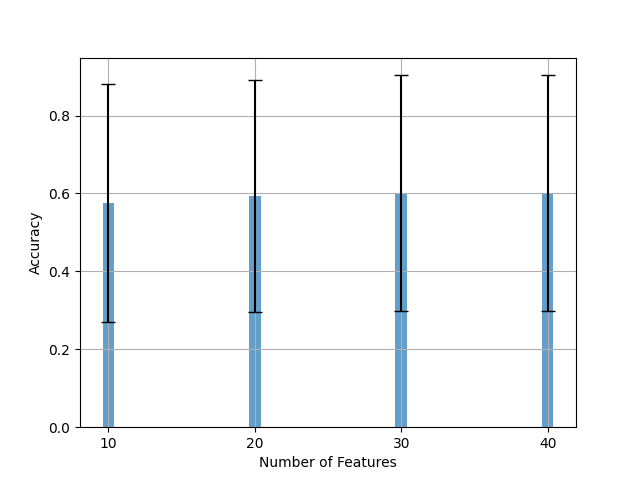}
  \caption{After augmentation.}
  \end{subfigure}
\caption{Test Accuracies of Random Forest (i=145)}
\label{fig:i145_rf_mfcc40_acc}
\end{figure}

As shown in table \ref{table:i145_rf_mfcc40_acc} and figure \ref{fig:i145_rf_mfcc40_acc}, the average accuracy of top 10 MFCC features is improved by 10.160\%, and the average accuracies of top 20, and 30 MFCC features are slightly improved after data augmentation. The average accuracy of 40 MFCC features is decreased by 3.118\% after data augmentation.
For the set 1, 2, 3, 4 with top 10 MFCC features, the accuracy increases from 43.921\% to 96.898\%, which is more than doubled.

\begin{table}[h!]
    \centering
    \refstepcounter{table}
    \textbf{\large Table \thetable: Test Accuracies of Random Forest (i=145)} \\[1ex]
    \begin{tabular}{c|c|c|c}
        \hline
        Training & Top & Before & After \\ [0.5ex]
        Set & Features & Aug. (\%) & Aug. (\%) \\ [0.5ex]
        \hline
        Average & 10 & 47.431 & 57.591 \\
        Average & 20 & 57.860 & 59.436 \\
        Average & 30 & 57.057 & 60.035 \\
        Average & 40 & 63.228 & 60.110 \\
        1,2,3,4 & 10 & 43.921 & 96.898 \\
        1,2,3,4 & 20 & 98.263 & 98.154 \\
        1,2,3,4 & 30 & 97.146 & 96.666 \\
        1,2,3,4 & 40 & 98.015 & 98.061 \\
        1,2,3,5 & 10 & 79.167 & 79.087 \\
        1,2,3,5 & 20 & 81.410 & 78.486 \\
        1,2,3,5 & 30 & 79.487 & 83.293 \\
        1,2,3,5 & 40 & 94.231 & 81.050 \\
        1,2,4,5 & 10 & 54.381 & 54.381 \\
        1,2,4,5 & 20 & 54.381 & 54.381 \\
        1,2,4,5 & 30 & 53.776 & 54.381 \\
        1,2,4,5 & 40 & 61.329 & 54.909 \\
        1,3,4,5 & 10 & 29.740 & 34.526 \\
        1,3,4,5 & 20 & 30.112 & 44.703 \\
        1,3,4,5 & 30 & 29.740 & 44.981 \\
        1,3,4,5 & 40 & 39.033 & 45.539 \\
        2,3,4,5 & 10 & 29.947 & 23.061 \\
        2,3,4,5 & 20 & 25.134 & 21.457 \\
        2,3,4,5 & 30 & 25.134 & 20.856 \\
        2,3,4,5 & 40 & 23.529 & 20.989 \\
        \hline
    \end{tabular}
    \caption*{The comparison of the test accuracies before and after data augmentation.}
    \label{table:i145_rf_mfcc40_acc}
\end{table}

\subsubsection{SVM RBF with 30 Instances}
\begin{figure}
  \begin{subfigure}{0.45\columnwidth}
  \includegraphics[width=\textwidth]{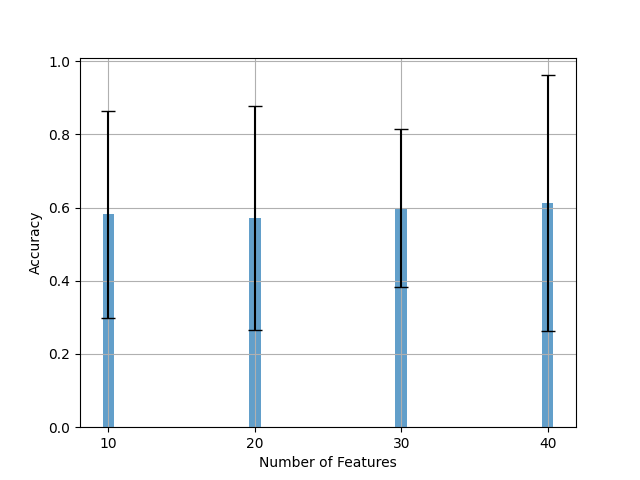}
  \caption{Before augmentation.}
  \end{subfigure}
  \hfill
  \begin{subfigure}{0.45\columnwidth}
  \includegraphics[width=\textwidth]{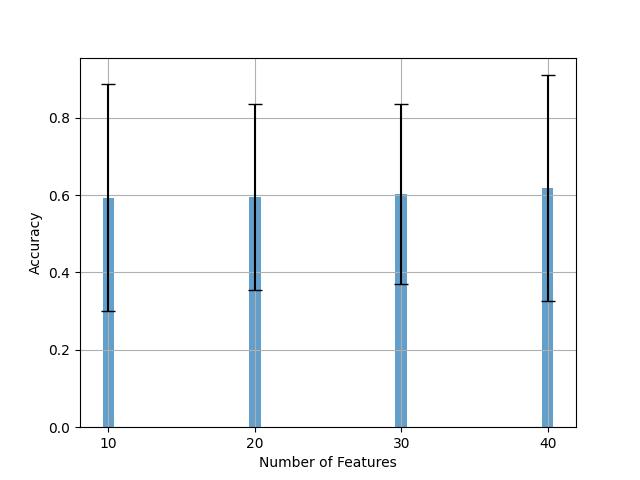}
  \caption{After augmentation.}
  \end{subfigure}
\caption{Test Accuracies of SVM RBF (i=30)}
\label{fig:i30_svmrbf_mfcc40_acc}
\end{figure}

Some accuracies before data augmentation are abnormally higher than others, such as the set 1, 2, 3, 4 with top 20 MFCC features and the set 1, 2, 3, 5 with 40 MFCC features have a 100\% accuracy, as shown in table \ref{table:i30_svmrbf_mfcc40_acc} and figure \ref{fig:i30_svmrbf_mfcc40_acc}. Overall, the average accuracies are improved slightly after data augmentation. The set 1, 2, 4, 5 remains the same after data augmentation, and the possible reasons can be: 1) the augmented instances are useless or redundant when using this training set, or 2) the SVM RBF classifier has already reached its performance upper bound with such instances.
In the future steps, we could try to use different data augmentation values, or try different original clips, to see what is the cause.

\begin{table}[h!]
    \centering
    \refstepcounter{table}
    \textbf{\large Table \thetable: Test Accuracies of SVM RBF (i=30)} \\[1ex]
    \begin{tabular}{c|c|c|c}
        \hline
        Training & Top & Before & After \\ [0.5ex]
        Set & Features & Aug. (\%) & Aug. (\%) \\ [0.5ex]
        \hline
        Average & 10 & 58.142 & 59.345 \\
        Average & 20 & 57.119 & 59.543 \\
        Average & 30 & 59.951 & 60.259 \\
        Average & 40 & 61.279 & 61.866 \\
        1,2,3,4 & 10 & 96.650 & 99.550 \\
        1,2,3,4 & 20 & 100.00 & 90.974 \\
        1,2,3,4 & 30 & 92.184 & 91.470 \\
        1,2,3,4 & 40 & 91.687 & 91.269 \\
        1,2,3,5 & 10 & 69.872 & 70.513 \\
        1,2,3,5 & 20 & 68.910 & 75.601 \\
        1,2,3,5 & 30 & 69.872 & 74.760 \\
        1,2,3,5 & 40 & 100.00 & 92.188 \\
        1,2,4,5 & 10 & 54.381 & 54.381 \\
        1,2,4,5 & 20 & 54.381 & 54.381 \\
        1,2,4,5 & 30 & 54.381 & 54.381 \\
        1,2,4,5 & 40 & 54.381 & 54.381 \\
        1,3,4,5 & 10 & 50.558 & 53.299 \\
        1,3,4,5 & 20 & 45.725 & 46.747 \\
        1,3,4,5 & 30 & 45.353 & 49.535 \\
        1,3,4,5 & 40 & 45.353 & 45.493 \\
        2,3,4,5 & 10 & 19.251 & 18.984 \\
        2,3,4,5 & 20 & 16.578 & 30.013 \\
        2,3,4,5 & 30 & 37.968 & 31.150 \\
        2,3,4,5 & 40 & 14.973 & 26.003 \\
        \hline
    \end{tabular}
    \caption*{The comparison of the test accuracies before and after data augmentation.}
    \label{table:i30_svmrbf_mfcc40_acc}
\end{table}

\subsubsection{SVM RBF with 145 Instances}
\begin{figure}
  \begin{subfigure}{0.45\columnwidth}
  \includegraphics[width=\textwidth]{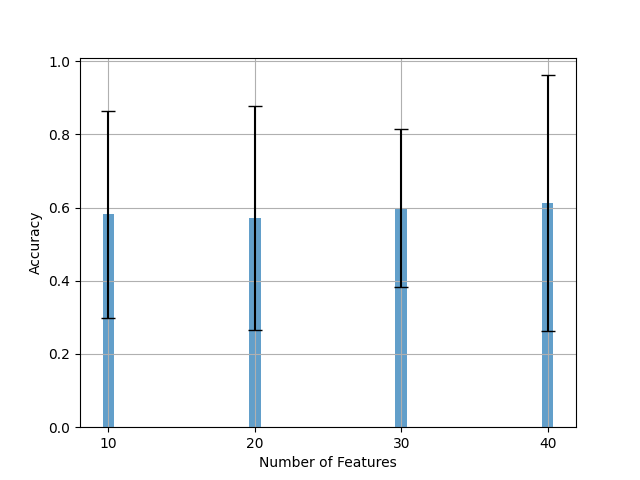}
  \caption{Before augmentation.}
  \end{subfigure}
  \hfill
  \begin{subfigure}{0.45\columnwidth}
  \includegraphics[width=\textwidth]{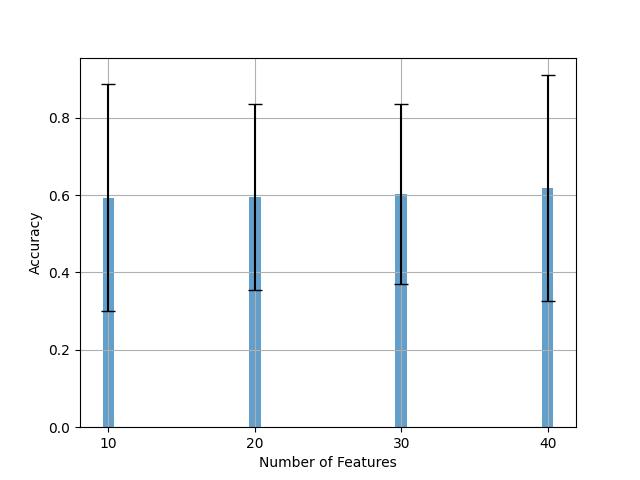}
  \caption{After augmentation.}
  \end{subfigure}
\caption{Test Accuracies of SVM RBF (i=30)}
\label{fig:i145_svmrbf_mfcc40_acc}
\end{figure}

As shown in table \ref{table:i145_svmrbf_mfcc40_acc} and figure \ref{fig:i145_svmrbf_mfcc40_acc}, SVM RBF with 145 instances have the exactly the same result as the one of SVM RBF with 30 instances. We have tried to retrain the model with same parameters (e.g., random state), but it gave the same output. One explanation could be that the additional instances are not significant so that no support vectors are introduced or changed when the SVM RBF model is fitting.

\begin{table}[h!]
    \centering
    \refstepcounter{table}
    \textbf{\large Table \thetable: Test Accuracies of SVM RBF (i=145)} \\[1ex]
    \begin{tabular}{c|c|c|c}
        \hline
        Training & Top & Before & After \\ [0.5ex]
        Set & Features & Aug. (\%) & Aug. (\%) \\ [0.5ex]
        \hline
        Average & 10 & 58.142 & 59.345 \\
        Average & 20 & 57.119 & 59.543 \\
        Average & 30 & 59.951 & 60.259 \\
        Average & 40 & 61.279 & 61.866 \\
        1,2,3,4 & 10 & 96.650 & 99.550 \\
        1,2,3,4 & 20 & 100.00 & 90.974 \\
        1,2,3,4 & 30 & 92.184 & 91.470 \\
        1,2,3,4 & 40 & 91.687 & 91.269 \\
        1,2,3,5 & 10 & 69.822 & 70.513 \\
        1,2,3,5 & 20 & 68.910 & 75.601 \\
        1,2,3,5 & 30 & 69.872 & 74.760 \\
        1,2,3,5 & 40 & 100.00 & 92.188 \\
        1,2,4,5 & 10 & 54.381 & 54.381 \\
        1,2,4,5 & 20 & 54.381 & 54.381 \\
        1,2,4,5 & 30 & 54.381 & 54.381 \\
        1,2,4,5 & 40 & 54.381 & 54.381 \\
        1,3,4,5 & 10 & 50.558 & 53.299 \\
        1,3,4,5 & 20 & 45.725 & 46.747 \\
        1,3,4,5 & 30 & 45.353 & 49.535 \\
        1,3,4,5 & 40 & 45.353 & 45.483 \\
        2,3,4,5 & 10 & 19.251 & 18.984 \\
        2,3,4,5 & 20 & 16.578 & 30.013 \\
        2,3,4,5 & 30 & 37.978 & 31.150 \\
        2,3,4,5 & 40 & 14.973 & 26.003 \\
        \hline
    \end{tabular}
    \caption*{The comparison of the test accuracies before and after data augmentation.}
    \label{table:i145_svmrbf_mfcc40_acc}
\end{table}

\subsubsection{XGBoost with 30 Instances}
\begin{figure}
  \begin{subfigure}{0.45\columnwidth}
  \includegraphics[width=\textwidth]{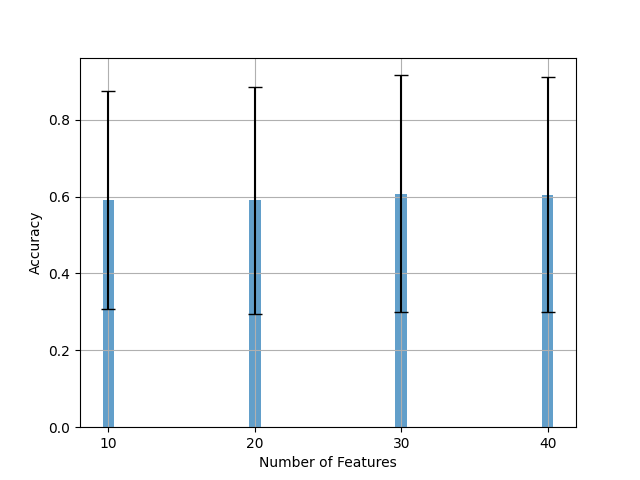}
  \caption{Before augmentation.}
  \end{subfigure}
  \hfill
  \begin{subfigure}{0.45\columnwidth}
  \includegraphics[width=\textwidth]{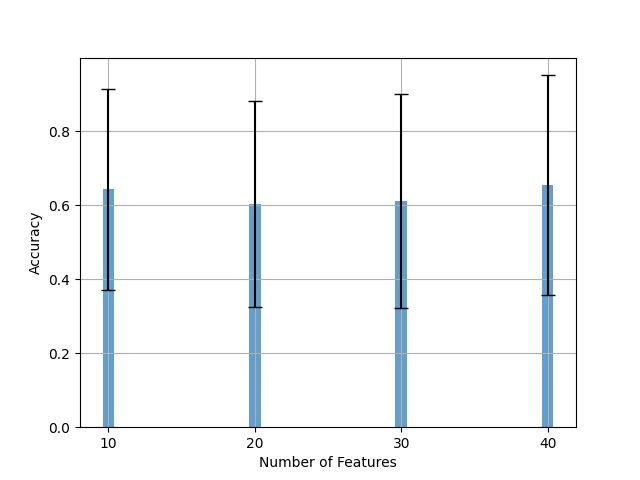}
  \caption{After augmentation.}
  \end{subfigure}
\caption{Test Accuracies of XGBoost (i=30)}
\label{fig:i30_xgboost_mfcc40_acc}
\end{figure}

As shown in figure \ref{table:i30_xgboost_mfcc40_acc} and figure \ref{fig:i30_xgboost_mfcc40_acc}, the average accuracies of top 10, 20, 30, 40 features are improved by 5.134\%, 1.363\%, 0.252\%, and 4.915\% after data augmentation, respectively. The set 1, 2, 3, 4 consistently has high accuracies before and after data augmentation. The largest improvement is set 1, 3, 4, 5 with top 10 MFCC features, and its accuracy improved from 57.249\% to 76.069\%. Conversely, the test accuracies of the set 2, 3, 4, 5 shows little to no improvement.

\begin{table}[h!]
    \centering
    \refstepcounter{table}
    \textbf{\large Table \thetable: Test Accuracies of XGBoost (i=30)} \\[1ex]
    \begin{tabular}{c|c|c|c}
        \hline
        Training & Top & Before & After \\ [0.5ex]
        Set & Features & Aug. (\%) & Aug. (\%) \\ [0.5ex]
        \hline
        Average & 10 & 59.090 & 64.224 \\
        Average & 20 & 59.005 & 60.368 \\
        Average & 30 & 60.808 & 61.060 \\
        Average & 40 & 60.531 & 65.446 \\
        1,2,3,4 & 10 & 98.263 & 95.037 \\
        1,2,3,4 & 20 & 96.278 & 96.386 \\
        1,2,3,4 & 30 & 97.643 & 98.325 \\
        1,2,3,4 & 40 & 99.752 & 97.519 \\
        1,2,3,5 & 10 & 73.718 & 72.716 \\
        1,2,3,5 & 20 & 78.205 & 76.923 \\
        1,2,3,5 & 30 & 83.974 & 78.846 \\
        1,2,3,5 & 40 & 79.808 & 81.851 \\
        1,2,4,5 & 10 & 41.088 & 54.305 \\
        1,2,4,5 & 20 & 40.785 & 54.381 \\
        1,2,4,5 & 30 & 42.296 & 54.343 \\
        1,2,4,5 & 40 & 41.692 & 54.192 \\
        1,3,4,5 & 10 & 57.249 & 76.069 \\
        1,3,4,5 & 20 & 58.364 & 51.487 \\
        1,3,4,5 & 30 & 58.736 & 51.394 \\
        1,3,4,5 & 40 & 59.480 & 73.281 \\
        2,3,4,5 & 10 & 25.134 & 22.995 \\
        2,3,4,5 & 20 & 21.390 & 22.660 \\
        2,3,4,5 & 30 & 21.390 & 22.393 \\
        2,3,4,5 & 40 & 21.925 & 20.388 \\
        \hline
    \end{tabular}
    \caption*{The comparison of the test accuracies before and after data augmentation.}
    \label{table:i30_xgboost_mfcc40_acc}
\end{table}

\subsubsection{XGBoost with 145 Instances}
\begin{figure}
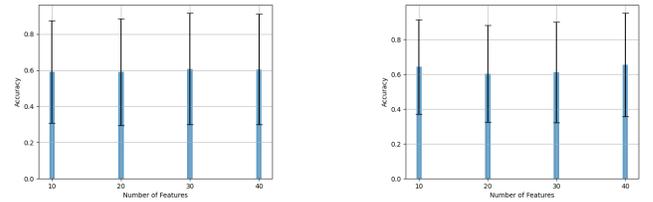

  \begin{subfigure}{0.45\columnwidth}
  \includegraphics[width=\textwidth]{figures/i=145_xgboost_mfcc40_1444338660_feature_selection.png}
  \caption{Before augmentation.}
  \end{subfigure}
  \hfill
  \begin{subfigure}{0.45\columnwidth}
  \includegraphics[width=\textwidth]{figures/i=145_xgboost_mfcc40_1444338660_augmented_feature_selection.png}
  \caption{After augmentation.}
  \end{subfigure}
\caption{Test Accuracies of XGBoost (i=145)}
\label{fig:i145_xgboost_mfcc40_acc}
\end{figure}

XGBoost with 145 instances have the exactly the same result as the one of XGBoost with 30 instances, as shown in table \ref{table:i145_xgboost_mfcc40_acc} and figure \ref{fig:i145_xgboost_mfcc40_acc}. A possible explanation could be that, since all instances are segmented from the same five original clips, they are very similar to each other, meaning that the other 115 instances would not provide useful information to the XGBoost model.

\begin{table}[h!]
    \centering
    \refstepcounter{table}
    \textbf{\large Table \thetable: Test Accuracies of XGBoost (i=145)} \\[1ex]
    \begin{tabular}{c|c|c|c}
        \hline
        Training & Top & Before & After \\ [0.5ex]
        Set & Features & Aug. (\%) & Aug. (\%) \\ [0.5ex]
        \hline
        Average & 10 & 59.090 & 64.224 \\
        Average & 20 & 59.005 & 60.368 \\
        Average & 30 & 60.808 & 61.060 \\
        Average & 40 & 60.531 & 65.446 \\
        1,2,3,4 & 10 & 98.263 & 95.037 \\
        1,2,3,4 & 20 & 96.278 & 96.386 \\
        1,2,3,4 & 30 & 97.643 & 98.325 \\
        1,2,3,4 & 40 & 99.752 & 97.519 \\
        1,2,3,5 & 10 & 73.718 & 72.716 \\
        1,2,3,5 & 20 & 78.205 & 76.923 \\
        1,2,3,5 & 30 & 83.974 & 78.846 \\
        1,2,3,5 & 40 & 79.808 & 81.851 \\
        1,2,4,5 & 10 & 41.088 & 54.305 \\
        1,2,4,5 & 20 & 40.785 & 54.381 \\
        1,2,4,5 & 30 & 42.296 & 54.343 \\
        1,2,4,5 & 40 & 41.692 & 54.192 \\
        1,3,4,5 & 10 & 57.249 & 76.069 \\
        1,3,4,5 & 20 & 58.364 & 51.487 \\
        1,3,4,5 & 30 & 58.736 & 51.394 \\
        1,3,4,5 & 40 & 59.480 & 73.281 \\
        2,3,4,5 & 10 & 25.134 & 22.995 \\
        2,3,4,5 & 20 & 21.390 & 22.660 \\
        2,3,4,5 & 30 & 21.390 & 22.393 \\
        2,3,4,5 & 40 & 21.925 & 20.388 \\
        \hline
    \end{tabular}
    \caption*{The comparison of the test accuracies before and after data augmentation.}
    \label{table:i145_xgboost_mfcc40_acc}
\end{table}

\subsubsection{k-NN with 30 Instances and 145 Instances}
As shown in table \ref{table:i30_knn_mfcc40_acc}, before data augmentation, the average accuracy increased with the number of features, reaching 62.651\% for the top 40 features. Among the training sets, 1, 2, 3, and 4 achieved the highest accuracy, with nearly perfect results of 99.876\% for the 40 MFCC features. Conversely, the set 2, 3, 4, and 5 has the lowest performance, with accuracies remaining below 30\% in all feature subsets.
The k-NN classifier has the same performance when using 145 instances instead of 30 instances.
The inability to handle augmented data with both 30 instances and 145 instances is likely because of the computational burden of k-NN's distance calculations. With an increased dataset size and high-dimensional features, k-NN's memory and computing requirements might exceed our system resources, eventually leading to kernel crashes. This highlights a critical limitation of k-NN for large-scale or augmented datasets, where its scalability becomes a challenge.

\begin{table}[h!]
    \centering
    \refstepcounter{table}
    \textbf{\large Table \thetable: Test Accuracies of k-NN (i=30)} \\[1ex]
    \begin{tabular}{c|c|c|c}
        \hline
        Training & Top & Before & After \\ [0.5ex]
        Set & Features & Aug. (\%) & Aug. (\%) \\ [0.5ex]
        \hline
        Average & 10 & 56.384 & - \\
        Average & 20 & 57.120 & - \\
        Average & 30 & 59.864 & - \\
        Average & 40 & 62.651 & - \\
        1,2,3,4 & 10 & 84.367 & - \\
        1,2,3,4 & 20 & 89.950 & - \\
        1,2,3,4 & 30 & 94.665 & - \\
        1,2,3,4 & 40 & 99.876 & - \\
        1,2,3,5 & 10 & 72.756 & - \\
        1,2,3,5 & 20 & 69.231 & - \\
        1,2,3,5 & 30 & 70.192 & - \\
        1,2,3,5 & 40 & 69.551 & - \\
        1,2,4,5 & 10 & 54.381 & - \\
        1,2,4,5 & 20 & 54.381 & - \\
        1,2,4,5 & 30 & 54.381 & - \\
        1,2,4,5 & 40 & 54.381 & - \\
        1,3,4,5 & 10 & 47.955 & - \\
        1,3,4,5 & 20 & 52.788 & - \\
        1,3,4,5 & 30 & 57.621 & - \\
        1,3,4,5 & 40 & 64.312 & - \\
        2,3,4,5 & 10 & 22.460 & - \\
        2,3,4,5 & 20 & 19.251 & - \\
        2,3,4,5 & 30 & 24.460 & - \\
        2,3,4,5 & 40 & 25.134 & - \\
        \hline
    \end{tabular}
    \caption*{The comparison of the test accuracies before and after data augmentation.}
    \label{table:i30_knn_mfcc40_acc}
\end{table}

\begin{figure}[h]
    \centering
    \includegraphics[width=0.45\textwidth]{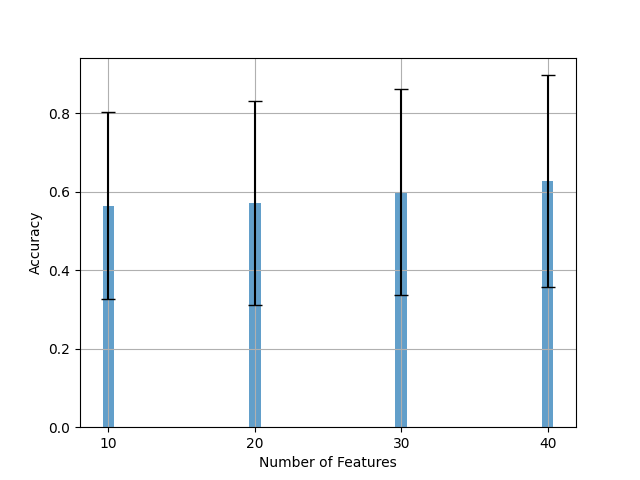}
    \caption{Test Accuracies of k-NN (i=30) before augmentation.}
    \label{fig:i30_knn_mfcc40_acc}
\end{figure}

\begin{table}[h!]
    \centering
    \refstepcounter{table}
    \textbf{\large Table \thetable: Test Accuracies of k-NN (i=145)} \\[1ex]
    \begin{tabular}{c|c|c|c}
        \hline
        Training & Top & Before & After \\ [0.5ex]
        Set & Features & Aug. (\%) & Aug. (\%) \\ [0.5ex]
        \hline
        Average & 10 & 56.384 & - \\
        Average & 20 & 57.120 & - \\
        Average & 30 & 59.864 & - \\
        Average & 40 & 62.651 & - \\
        1,2,3,4 & 10 & 84.367 & - \\
        1,2,3,4 & 20 & 89.950 & - \\
        1,2,3,4 & 30 & 94.665 & - \\
        1,2,3,4 & 40 & 99.876 & - \\
        1,2,3,5 & 10 & 72.756 & - \\
        1,2,3,5 & 20 & 69.231 & - \\
        1,2,3,5 & 30 & 70.192 & - \\
        1,2,3,5 & 40 & 69.551 & - \\
        1,2,4,5 & 10 & 54.381 & - \\
        1,2,4,5 & 20 & 54.381 & - \\
        1,2,4,5 & 30 & 54.381 & - \\
        1,2,4,5 & 40 & 54.381 & - \\
        1,3,4,5 & 10 & 47.955 & - \\
        1,3,4,5 & 20 & 52.788 & - \\
        1,3,4,5 & 30 & 57.621 & - \\
        1,3,4,5 & 40 & 64.312 & - \\
        2,3,4,5 & 10 & 22.460 & - \\
        2,3,4,5 & 20 & 19.251 & - \\
        2,3,4,5 & 30 & 22.460 & - \\
        2,3,4,5 & 40 & 25.134 & - \\
        \hline
    \end{tabular}
    \caption*{The comparison of the test accuracies before and after data augmentation.}
    \label{table:i145_knn_mfcc40_acc}
\end{table}

\begin{figure}[h]
    \centering
    \includegraphics[width=0.45\textwidth]{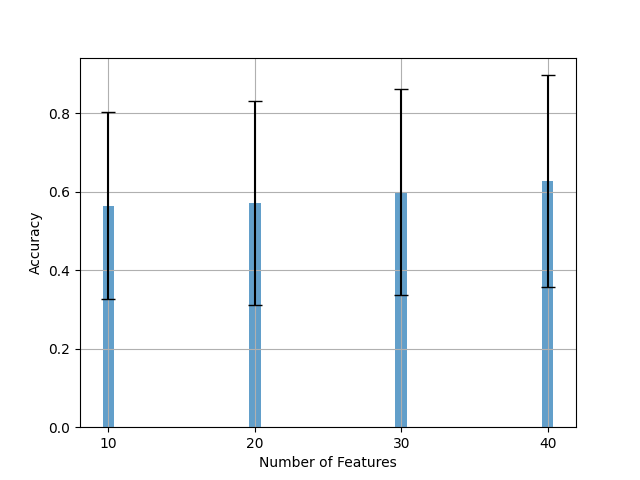}
    \caption{Test Accuracies of k-NN (i=145) before augmentation.}
    \label{fig:i145_knn_mfcc40_acc}
\end{figure}

\subsubsection{Explanation and Summary}
\begin{figure}[h]
    \centering
    \includegraphics[width=0.45\textwidth]{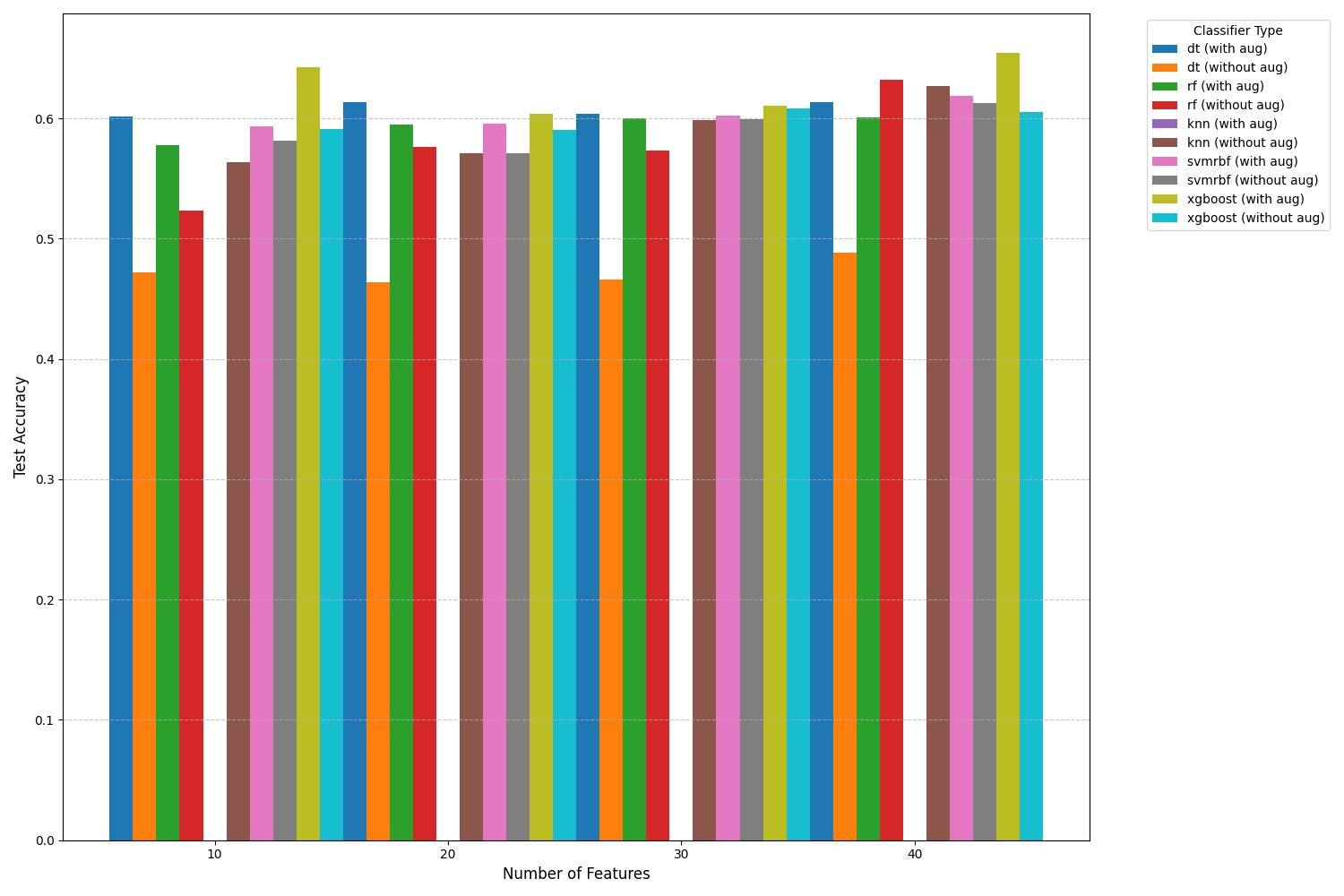}
    \caption{Test Accuracies from All Classifiers.}
    \label{fig:test_acc_from_all}
\end{figure}

Data augmentation significantly improves test accuracies for most classifiers, especially Decision Tree and Random Forest. It slightly improves test accuracies for SVM RBF and XGBoost. Figure \ref{fig:test_acc_from_all} shows the average test accuracies from all trained classifiers.

\begin{itemize}
    \item Decision Tree: It is benefited greatly from data augmentation, showing consistent improvements in test accuracies.
    \item Random Forest: Before data augmentation, its performance decreases when using a larger training set. After data augmentation, both results for 30 instances and 145 instances are similar.
    \item SVM RBF: It shows slight accuracy improvement after data augmentation, while certain training sets cannot be improved and has exactly the same results before and after augmentation.
    \item XGBoost: It has the highest average test accuracies after data augmentation, that it is above 65\% when using all 40 MFCC features and is about 64\% when using top 10 MFCC features, as shown in figure \ref{fig:test_acc_from_all}.
    \item k-NN: It receives moderate test accuracies compared to other classifiers before data augmentation. We failed to test its performance after data augmentation due to limited computing resources.
\end{itemize}

Increasing the number of MFCC features generally leads to higher accuracies across classifiers. When using all 40 MFCC features, it would have the best performance.

Training set 1, 2, 3, 4 produces the best test accuracies, which are often exceeding 90\%. In contrast, training set 2, 3, 4, 5 consistently produces lowest test accuracies among all training sets.

\section{Discussion}
The result of this study illustrate the effectiveness of using acoustic features, such as MFCC, and machine learning models, including decision tree, random forest, k-NN, SVM RBF, and XGBoost, for insect sound classification.

This study has some limitations:
\begin{itemize}
    \item Since we use data augmentation to generate more instances to train and test, the classifiers have the possibility to over-fit.
    \item It is hard to manually segment insect sounds accurately and precisely, because many sounds are extremely short and have to be discarded from the training set. Besides, many original clips in the original datasets have huge background noises, that could potentially affect the accuracy dramatically.
    \item As shown in the t-SNE and UMAP plots, many instances from different classes are close to each other and some original clips in the same class are far apart from each other, indicating that it could be hard for machine learning models to classify among them.
\end{itemize}

Future work could focus on the following points to further contribute to this area:
\begin{itemize}
    \item Currently we create a small dataset based on several other datasets, and eventually only 5 original clips are used for each class among all datasets. In the future work, we could increase the number of the original clips, so that we will have more class balanced instances before data augmentation.
    \item In this study, we use window size 0.1 seconds to segment original clips into instances. In the future work, we could try different window size, such as 0.01 seconds, 1.0 seconds, and other values, to test the performances and trends when using different window sizes.
    \item In this study, we only train and test the classifiers based on the existing dataset. In the future work, we should test all trained classifiers in the real world, to verify their performances and usefulness.
\end{itemize}

\bibliographystyle{IEEEtran}
\bibliography{References}

\end{document}